\documentclass[aps,pre,reprint,showpacs,bibnotes,twoside]{revtex4-1}
\usepackage{epsfig}
\usepackage{color}
\usepackage{bm}
\usepackage{latexsym}
\usepackage{amssymb}
\usepackage{amsmath}
\usepackage{float}
\usepackage{graphicx}
\usepackage{mathtools}
\usepackage{epstopdf}
\usepackage{tikz}
\usepackage{etoolbox}
\usepackage{tikz}
\usepackage{times}

\newrobustcmd*{\mycircle}[1]{\tikz{\filldraw[draw=#1,fill=#1] (0,0) circle [radius=0.05cm];}}
\newrobustcmd*{\mydiamond}[1]{
    \tikz{
        \filldraw[draw=#1,fill=#1, rotate around={45:(0,0)}] 
        (0,0)
        rectangle
        (0.1cm,0.1cm);
    }
}

\begin{document}

\newcommand{\iu}{\mathrm{i}}
\newcommand{\hide}[1]{}
\newcommand{\tbox}[1]{\mbox{\tiny #1}}
\newcommand{\half}{\mbox{\small $\frac{1}{2}$}}
\newcommand{\sinc}{\mbox{sinc}}
\newcommand{\const}{\mbox{const}}
\newcommand{\trc}{\mbox{trace}}
\newcommand{\intt}{\int\int }
\newcommand{\ointt}{\int\int\circ\ }
\newcommand{\eexp}{\mbox{e}^}
\newcommand{\bra}{\left\langle}
\newcommand{\ket}{\right\rangle}
\newcommand{\EPS} {\mbox{\LARGE $\epsilon$}}
\newcommand{\ar}{\mathsf r}
\newcommand{\im}{\mbox{Im}}
\newcommand{\re}{\mbox{Re}}
\newcommand{\bmsf}[1]{\bm{\mathsf{#1}}}
\newcommand{\mpg}[2][1.0\hsize]{\begin{minipage}[b]{#1}{#2}\end{minipage}}

\title{Spacing ratio characterization of the spectra of directed random networks}

\author{Thomas Peron,$^1$ Bruno Messias F. de Resende,$^2$ Francisco A. Rodrigues,$^1$ Luciano da F. Costa$^2$, J. A. M\'endez-Berm\'udez,$^{1,3}$}

\affiliation{
$^1$Institute of Mathematics and Computer Science, University of S\~{a}o Paulo, S\~{a}o Carlos 13566-590, S\~{a}o Paulo, Brazil\\
$^2$S\~{a}o Carlos Institute of Physics, University of S\~{a}o Paulo, S\~{a}o Carlos, S\~{a}o Paulo, Brazil \\
$^3$Instituto de F\'isica, Benem\'erita Universidad Aut\'onoma de Puebla, Apartado postal J-48, Puebla 72570, M\'exico }

\date{\today}

\begin{abstract}
Previous literature on random matrix and network science has traditionally employed measures derived from nearest-neighbor level spacing distributions 
to characterize the eigenvalue statistics of random matrices. This approach, however, depends crucially 
on eigenvalue unfolding procedures, which in many situations represent a major hindrance due to constraints in the calculation, specially in the case of complex spectra. Here we study the spectra of directed networks using the recently introduced ratios between nearest- and next-to-nearest eigenvalue spacing, thus circumventing the shortcomings imposed by spectral unfolding. Specifically, we characterize the eigenvalue 
statistics of directed Erd\H{o}s-R\'enyi (ER) random networks by means of two adjacency matrix representations; namely (i) weighted non-Hermitian random matrices and (ii) a transformation on non-Hermitian adjacency matrices which produces weighted Hermitian matrices. For both representations, we find that the distribution of spacing ratios becomes universal for a fixed average degree, in accordance with undirected random networks. Furthermore, by calculating the average spacing ratio as a function of the average degree, we show that the spectral statistics of directed ER random networks undergoes a transition from Poisson to Ginibre statistics for model (i) and from Poisson to Gaussian Unitary Ensemble statistics for model (ii). Eigenvector delocalization effects of directed networks are also discussed.
\end{abstract}


\maketitle



\section{Introduction}

Networks have become crucial tools for the modeling of different types of complex systems composed of discrete units. Prominent examples include technological systems, as in the case of the World Wide Web (WWW)~\cite{newman2018networks}, Internet~\cite{newman2018networks}, and power-grids~\cite{pagani2013power,arianos2009power}; social networks, both off- and on-line~\cite{newman2018networks}; biological systems, like foods webs~\cite{newman2018networks,allesina2008general} and mutualistic relationships between species~\cite{bascompte2013mutualistic};  and many others~\cite{newman2018networks}. A substantial part of these networks are said to be directed, in the sense that interactions between its components occur asymmetrically; that is, using the WWW as example, there may be links from one page to others, but not necessarily links pointing back. 

The advances in the characterization of the structure of networks have also improved our understanding about the functioning of the systems they represent. In particular, the performance of several dynamical processes (such as, epidemic spreading, synchronization, and percolation) can, in general, be quantified in terms of spectral properties of adjacency matrices, which in turn encode the network topology~\cite{porter2016dynamical}. Progress in this area, however, has been mainly concentrated on the dynamics of random undirected networks, i.e., networks that are characterized by sparse Hermitian random matrices and to which several results obtained in Random Matrix Theory (RMT) are applicable~\cite{M04}. 

Despite the importance of complex systems whose interactions are asymmetric, spectral properties of directed networks have been much less explored than their undirected counterparts. The reason for this might reside in the difficulty of adapting analytical techniques developed for Hermitian matrices to the analysis of the complex spectra of sparse non-Hermitian ones. Indeed, only very recently rigorous calculations have started to be obtained for the spectral density of sparse non-Hermitian matrices (see, e.g.,~\cite{fyodorov1997almost,rogers2009cavity,neri2012spectra,saade2014spectral,metz2019spectral}). Furthermore, results concerning the universality of spectral features of such matrices are even scarcer when compared to the corresponding literature on random matrices derived from undirected random graphs~\cite{metz2019spectral}. 

Besides being interesting in its own right, the identification of universality classes in spectral properties can also be relevant to the study of dynamical processes running on directed networks: by detecting the spectral observables that remain independent from details of the random matrix realization, one is able to infer what global network properties control dynamical transitions in the complex system under study; examples of the application of universal spectral properties are found in the stability criteria of large ecosystems~\cite{may1972will,allesina2012stability} and other processes on directed networks~\cite{metz2019spectral}. Motivated by these facts, in this paper, we carry out an extensive analysis of the spectral properties of sparse Hermitian and non-Hermitian matrices, both representing directed random networks.

Certainly, the most popular tool used to characterize the spectral properties of random matrix 
ensembles has been the nearest-neighbor energy-level spacing distribution $P(s)$~\cite{M04}.
It was originally defined for real spectra~\cite{M04} and later also extended to complex spectra~\cite{GHS88,MPW99}. 
However, the computation of $P(s)$ from complex spectra remains a subject to be further developed. We believe that this may be due to the problem of spectrum unfolding that, even for real 
spectra, may become a cumbersome task; see e.g.~\cite{GMRR2002,Abuelenin2012,Abuelenin2018}. Spectrum unfolding, in random matrix theory (RMT), 
is the process of locally normalizing a spectrum such that the mean level spacing $\left< s \right>$ equals unity.
Fortunately, recently, the problem of spectrum unfolding has already been circumvented, for real spectra, 
by the introduction of the distribution of the ratio between consecutive level spacings $P(r)$~\cite{OH07,ABG13}.
Moreover, very recently, the version of $P(r)$ for complex spectra was proposed in Ref.~\cite{SRP20}.

In this paper, we employ real and complex spacing ratios in order to characterize the spectral properties of directed networks. We address this task by considering two adjacency matrix representations of Erd\H{os}-R\`enyi (ER) random networks; namely, weighted non-Hermitian adjacency matrices and a recently introduced operator~\cite{guo2017hermitian,liu2015hermitian} which yields complex Hermitian adjacency matrices (see Sec.~\ref{sec:models_quantities} for definitions). Therefore, since here we are dealing with real and complex spectra (i.e.~Hermitian and non-Hermitian 
matrices) we shall compute both real and 
complex versions of $P(r)$. More precisely, we will concentrate on the average ratio $\left< r \right>$ 
as a complexity indicator to characterize
the localization-to-delocalization transition of the random matrix models we will use as representations 
of directed random networks. It is relevant to stress that due to the need of spectral unfolding,
the use of $\left< s \right>$ as complexity indicator is not feasible due to the constraint of having $\left< s \right>=\mbox{const.}=1$ after unfolding; for this reason, we rely our analysis on the characterization of $\langle r \rangle$ as a function of the global network parameters, such as number of nodes and average degree.  



\section{Models and quantities}
\label{sec:models_quantities}
\subsection{Models}

We consider directed random networks $G$ from the standard ER model $G(n,p)$, i.e., 
$G$ has $n$ vertices and each directed edge appears independently with probability $p \in (0,1)$. Given a directed network $G(n,p)$ we analyze the spectral properties of two different matrix 
representations:

(i) {\it The randomly-weighted non-Hermitian adjacency matrix} $\mathbf{A}_{\tbox{dRGE}}$.

The matrix $\mathbf{A}_{\tbox{dRGE}}$ is constructed as follows: a random directed ER graph 
is constructed and its adjacency matrix is extracted, then the adjacency 
matrix
is weighted with random variables (including self loops). Thus we get the matrix:
\begin{equation}
[\mathbf{A}_{\tbox{dRGE}}]_{uv}=\left\{
\begin{array}{ll}
\epsilon_{uu} & \mbox{if $u=v$}, \\
\epsilon_{uv} & \mbox{if $u\rightarrow v$}, \\
0 & \mbox{otherwise},
\end{array}
\right.
\label{AdRGE}
\end{equation}
where $u \rightarrow v$ denotes that there exists a directed edge from node $u$ to $v$. Here, we choose $\epsilon_{uv}$ as statistically-independent random variables drawn from a 
normal distribution with zero mean and variance one, $\epsilon_{uv}\sim\mathcal {N}(0, 1)$. Evidently, since $G$ is directed, 
$\epsilon_{uv}\ne \epsilon_{vu}$; thus, matrix $\mathbf{A}_{\tbox{dRGE}}$ is non-Hermitian.
We use the subscript ``dRGE'' because we identify the matrix $\mathbf{A}_{\tbox{dRGE}}$ as a
{\it diluted} version of the Real Ginibre Ensemble (RGE)~\cite{G65}; i.e.~for a complete network, 
when $p=1$, $\mathbf{A}_{\tbox{dRGE}}$ is a member of the RGE (the RGE consists of random 
$n\times n$ matrices formed from independent and identically distributed standard Gaussian entries); some spectral properties of the RGE were reported in~\cite{FN07}. 
Also note that when $p=0$,
for a completely disconnected network, $\mathbf{A}_{\tbox{dRGE}}$ reproduces the Poisson
Ensemble (PE)~\cite{M04}; that is,~$\mathbf{A}_{\tbox{dRGE}}$ becomes a diagonal random matrix.
Thus, a transition from the PE to the RGE is expected when increasing $p$ from zero to one.


(ii) {\it The randomly-weighted Hermitian adjacency matrix} $\mathbf{A}_{\tbox{M}}$.

Recently, a Hermitian adjacency operator for unweighted directed graphs was defined in Refs.~\cite{guo2017hermitian,liu2015hermitian}. Interestingly, it turns out that the adjacency operator of Refs.~\cite{guo2017hermitian,liu2015hermitian} is a special case of a more generic one originated from the magnetic Laplacian formalism~\cite{lieb1993,berkolaiko2013,fanuel2017magnetic}; see Appendix~\ref{appendixMoharMagnetic} for more details. Given  that equivalence we call the Hermitian adjacency matrix associated with a directed network just as the {\it magnetic adjacency matrix}.
Owing to the numerous recent applications of the magnetic Laplacian formalism, we choose to study the properties of a random ensemble associated with it. The magnetic random ensemble is created with the following steps: a random directed ER graph is created; the magnetic adjacency matrix is thereby extracted from the graph and is then weighted with random variables. By denoting by $\mathbf{A}_{\tbox{ER}}$ the binary adjacency matrix extracted a from directed ER graph, this procedure, therefore, gives us the following random matrix:
\begin{equation}
[\mathbf{A}_{\tbox{M}}]_{uv}=\left\{
\begin{array}{ll}
\epsilon_{uu} & \mbox{if $u=v$}, \\
\epsilon_{uv} [\mathbf{A}_{\tbox{ER}}]_{vu} & \mbox{if $u\leftrightarrow v$}, \\
-\imath \epsilon_{uv}[\mathbf{A}_{\tbox{ER}}]_{vu} & \mbox{if $u\rightarrow v$}, \\
0 & \mbox{otherwise}.
\end{array}
\right.
\label{AH}
\end{equation}
Again, we choose $\epsilon_{uv}$ as statistically-independent random variables drawn from a 
normal distribution with zero mean and variance one. Indeed, 
$[\mathbf{A}_{\tbox{M}}]_{vu}=[\mathbf{A}_{\tbox{M}}]_{uv}^*$ by 
construction. In this case, for increasing $p$, the ensemble defined by $\mathbf{A}_{\tbox{M}}$ transits from the 
PE, when $p=0$, to real symmetric full random matrices, when $p=1$. The later ensemble is very 
similar to the Gaussian Orthogonal Ensemble~\cite{M04} of RMT, but not exactly equal; in the GOE the
diagonal matrix elements have twice the variance than the off-diagonal ones.

\subsection{Quantities}



Below we follow a recently introduced approach under which the adjacency matrices of random 
graphs are studied statistically. See the application of this approach on undirected ER 
graphs~\cite{MAM15,GAC18,MM19,TFM19,MMRS20}, random regular and random rectangular 
graphs~\cite{AMGM18}, $\beta$-skeleton graphs~\cite{AME19}, multiplex and multilayer 
networks~\cite{MFMR17}, and bipartite graphs~\cite{MAMPS19}.

In the next Section we characterize the real spectra of $\mathbf{A}_{\tbox{M}}$ and 
the complex spectra of $\mathbf{A}_{\tbox{dRGE}}$ by computing, respectively, the average values of
the ratio between consecutive level spacings $r_\mathbb{R}$ and the ratio between 
nearest- and next-to-nearest neighbor spacings $r_\mathbb{C}$, which are defined 
as follows.
On the one hand, given the real ordered spectrum 
$\lambda_1>\lambda_2>\cdots>\lambda_{n-1}>\lambda_n$, the $k$-th ratio $r_\mathbb{R}^k$ 
reads as~\cite{OH07,ABG13}
\begin{equation}
\label{rR}
r_\mathbb{R}^k = \frac{\min(\lambda_{k+1}- \lambda_k,\lambda_{k}- \lambda_{k-1})}{\max(\lambda_{k+1}- \lambda_k,\lambda_{k}- \lambda_{k-1})} \ .
\end{equation}
Here, $r_\mathbb{R}\in[0,1]$.
On the other hand, given the complex spectrum $\{ \lambda_k \}$ the $k$-th ratio $r_\mathbb{C}^k$ 
reads as~\cite{SRP20}
\begin{equation}
r_\mathbb{C}^k = \frac{\left| \lambda^{NN}_k - \lambda_k \right|}{\left| \lambda^{NNN}_k - \lambda_k \right|} \ ,
\label{rC}
\end{equation}
where $\lambda^{NN}_k$ and $\lambda^{NNN}_k$ are, respectively, the nearest and the next-to-nearest 
neighbors of $\lambda_k$ in $\mathbb{C}$. Note that, as well as $r_\mathbb{R}$, $r_\mathbb{C}\in[0,1]$. Moreover, note that $r_\mathbb{C}$ can also be computed for real spectra.

\section{Results}

Now we use exact numerical diagonalization to obtain the eigenvalues $\lambda_k$ ($k =1,\ldots,n$) 
of large ensembles of matrices given by Eqs.~(\ref{AdRGE}) and~(\ref{AH}) (characterized by $n$ 
and $p$) and compute the average values of the ratios $r_\mathbb{C}$ and $r_\mathbb{R}$.

\begin{figure*}[!t]
\includegraphics[width=0.8\textwidth]{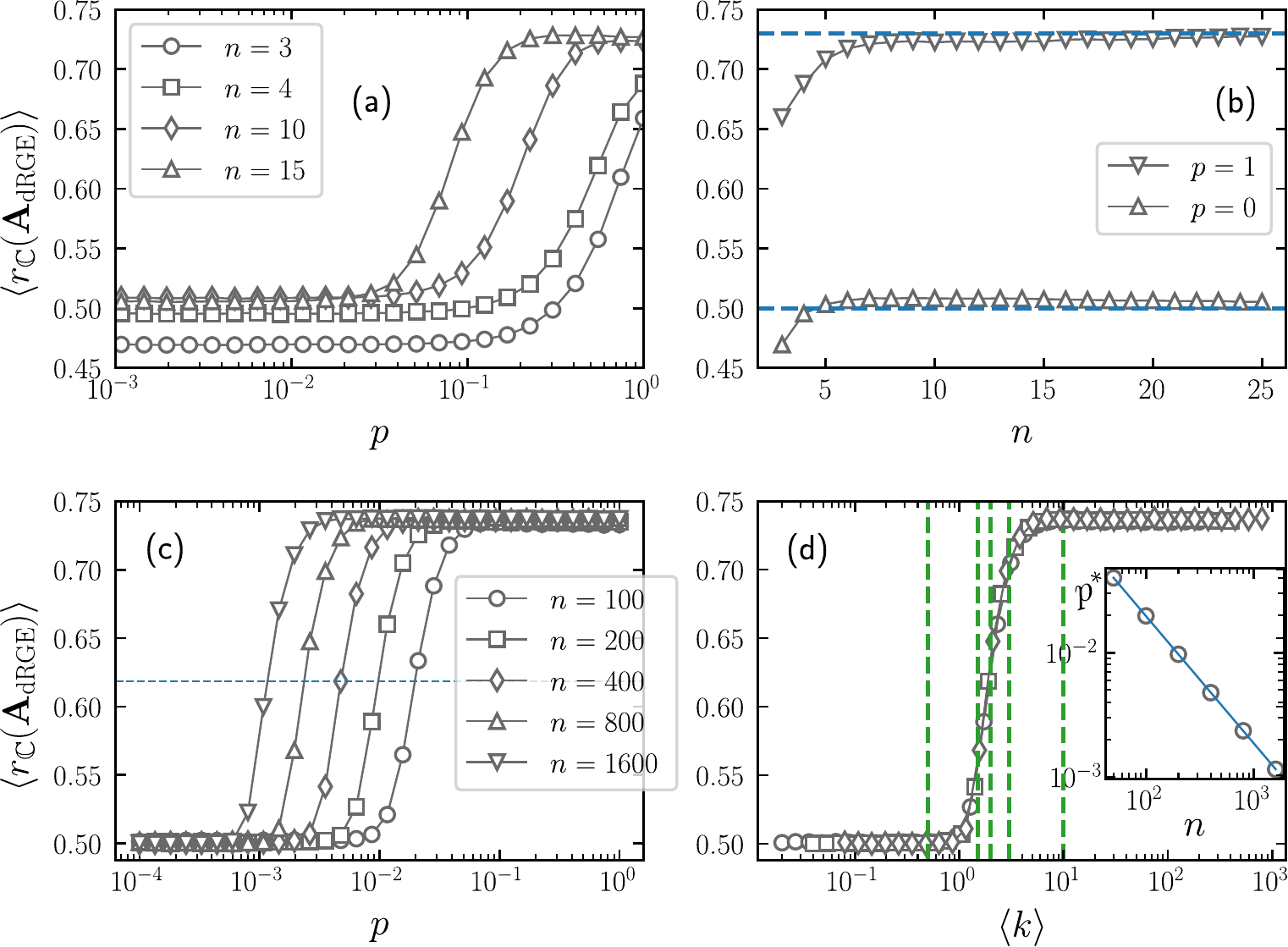}
\caption{
(a,c) Ensemble average of the ratio $r_\mathbb{C}$ for the adjacency matrices represented by the 
diluted real Ginibre ensemble, $\left< r_\mathbb{C}(\mathbf{A}_{\tbox{dRGE}}) \right>$, as a function 
of the probability $p$ for several network sizes $n$.
The horizontal dashed line in panel (c) indicates 
$\left< r_\mathbb{C}(\mathbf{A}_{\tbox{dRGE}}) \right>=0.6188$.
(b) $\left< r_\mathbb{C}(\mathbf{A}_{\tbox{dRGE}}) \right>$ as a function of $n$ for $p=1$ and $p=0$. 
The horizontal dashed lines at $\left< r_\mathbb{C}(\mathbf{A}_{\tbox{dRGE}}) \right>=0.7370$ and 
0.5006 indicate the values of $\left< r_\mathbb{C}(\mathbf{A}_{\tbox{dRGE}}) \right>$ for $p=1$ and 
$p=0$, respectively, at $n=1000$.
(d) Same curves of panel (c) but as a function of the average degree $\left< k \right>$.
Vertical dashed lines mark the values of $\left< k \right>$ (0.5, 1.5, 2, 3, and 10) chosen to report
the PDFs of $r_\mathbb{C}$, $P(r_\mathbb{C})$, in Fig.~\ref{Fig02}.
Inset: $p^*$ as a function of $n$. The dashed line is the fitting of Eq.~(\ref{scaling}) to the the data 
with fitting parameters $\mathcal{C} = 2.1949$ and $\delta = -1.0235$.
Each symbol was computed from the ratios of $10^6/n$ directed random networks $G(n,p)$.}
\label{Fig01}
\end{figure*} 

\subsection{Diluted real Ginibre ensemble}
\label{sec:diluted_real_ginibre_ensemble}

In Fig.~\ref{Fig01}(a,c) we present the average of the ratio $r_\mathbb{C}$ for the adjacency matrices 
represented by the diluted real Ginibre ensemble, $\left< r_\mathbb{C}(\mathbf{A}_{\tbox{dRGE}}) \right>$, 
as a function of the probability $p$ for several network sizes $n$.
All averages here and below are computed from the ratios of $10^6/n$ directed random networks $G(n,p)$.
We observe that the curves of $\left< r_\mathbb{C}(\mathbf{A}_{\tbox{dRGE}}) \right>$, for $n\ge 50$, 
have a very similar shape as a function of $p$: $\left< r_\mathbb{C}(\mathbf{A}_{\tbox{dRGE}}) \right>$ 
shows a smooth transition (in log scale) from $\approx 0.5$ to $\approx 0.737$ when $p$ increases from 
zero (isolated vertices) to one (complete networks); see Fig.~\ref{Fig01}(c). 
For smaller network sizes, $n<50$, clear small-size effects appear, as can be seen in Fig.~\ref{Fig01}(a).
Indeed, in Fig.~\ref{Fig01}(b) we show the small size dependence of 
$\left< r_\mathbb{C}(\mathbf{A}_{\tbox{dRGE}}) \right>$ for the two limiting values of $p$: zero and one.

\begin{figure*}[!t]
\includegraphics[width=1.0\textwidth]{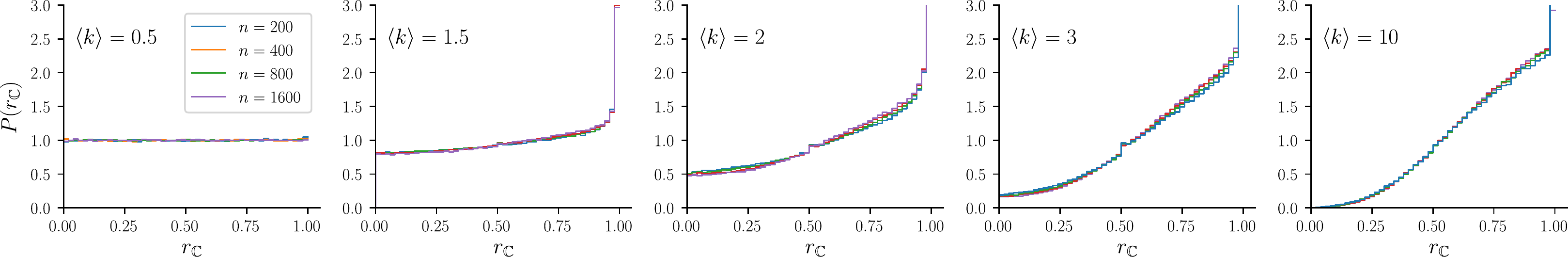}
\caption{
Probability density function of the ratio $r_\mathbb{C}$, $P(r_\mathbb{C})$, for the adjacency 
matrix represented by the diluted real Ginibre ensemble. Each panel, corresponding to different values 
of the average degree $\left< k \right>$, contains histograms of four different network sizes $n$.
The values of $\left< k \right>=0.5$, 1.5, 2, 3, and 10 are marked as vertical dashed lines in Fig.~\ref{Fig01}(d).
Each histogram was constructed from the ratios of $10^6/n$ directed random networks $G(n,p)$.}
\label{Fig02}
\end{figure*}

From Fig.~\ref{Fig01}(c) we can clearly see that the main effect of increasing $n$ is the 
displacement of the curves $\left< r_\mathbb{C}(\mathbf{A}_{\tbox{dRGE}}) \right>$ vs.~$p$ to the left on 
the $p$-axis. Moreover, the fact that these curves, plotted in semi-log scale, are shifted the same amount 
on the $p$-axis when doubling $n$ make us anticipate the existence of a scaling parameter that depends 
on $n$. In order to search for that scaling parameter we first establish a measure to characterize the 
position of the curves $\left< r_\mathbb{C}(\mathbf{A}_{\tbox{dRGE}}) \right>$ on the $p$-axis: We 
choose the value of $p$, that we label as $p^*$, for which 
$\left< r_\mathbb{C}(\mathbf{A}_{\tbox{dRGE}}) \right>$ approaches half of the full transition, see the horizontal dashed line in Fig.~\ref{Fig01}(c) at $\left< r_\mathbb{C}(\mathbf{A}_{\tbox{dRGE}}) \right>=0.6188$. Notice that $p^*$ characterizes the transition from isolated 
vertices to complete networks of size $n$.

Then, in the inset of Fig.~\ref{Fig01}(d) we plot $p^*$ versus $n$. The linear trend of the data (in log-log 
scale) suggests the power-law behavior
\begin{equation}
\label{scaling}
p^* = \mathcal{C} n^\delta .
\end{equation}
In fact, Eq.~(\ref{scaling}) provides an excellent fitting to the data with $\delta \approx -1$. 
Therefore, by plotting again the curves of $\left< r_\mathbb{C}(\mathbf{A}_{\tbox{dRGE}}) \right>$ 
now as a function of the probability $p$ divided by $p^*$, 
\begin{equation}
\label{k}
\frac{p}{p^*} \propto \frac{p}{n^\delta} \approx \frac{p}{n^{-1}} = np \equiv \left< k \right> \ ,
\end{equation} 
we observe that curves for different graph sizes $n$ collapse on top of a single {\it universal} curve, 
see Fig.~\ref{Fig01}(d). This means that once the average degree $\left< k \right>$ is fixed, the 
average ratio 
$r_\mathbb{C}(\mathbf{A}_{\tbox{dRGE}})$ of the diluted RGE is also fixed. This statement is in 
accordance with the results reported in~\cite{MMRS20,MAM15,MAM13}, where topological, spectral 
and transport properties of undirected ER graphs where shown to be universal for the 
product $np$, see also~\cite{GAC18,MM19,TFM19}.

Notice that Fig.~\ref{Fig01}(d) provides a way to identify the statistical regimes of  
$\left< r_\mathbb{C}(\mathbf{A}_{\tbox{dRGE}}) \right>$ once the average degree $\left< k \right>$
is known:
When $\left< k \right><1$, 
$\left< r_\mathbb{C}(\mathbf{A}_{\tbox{dRGE}}) \right>=\left< r_\mathbb{C}(\mbox{PE}) \right>\approx 0.5$;
i.e.~the value of $\left< r_\mathbb{C} \right>$ corresponding to the PE.
For $\left< k \right>>7$, $\left< r_\mathbb{C}(\mathbf{A}_{\tbox{dRGE}}) \right>=\left< r_\mathbb{C}(\mbox{RGE}) \right>\approx 0.737$; that is, the value of $\left< r_\mathbb{C} \right>$ corresponding to the RGE.
While the transition region is defined for $1<\left< k \right><7$.
Thus, $\left< k \right>=1$ and 7 mark the onset of the delocalization transition and the onset of the RGE 
limit, respectively.

Now in Fig.~\ref{Fig02} we show PDFs of the ratio $r_\mathbb{C}$, $P(r_\mathbb{C})$, for selected 
values of $\left< k \right>$ (marked as vertical dashed lines in Fig.~\ref{Fig01}(d)). 
Each panel of Fig.~\ref{Fig02} contains histograms of four different network sizes $n$ that fall one on 
top of the other, except for small size effects visible mainly in the transition region
$1<\left< k \right><7$. With this, we validate that the invariance of the average of 
$r_\mathbb{C}(\mathbf{A}_{\tbox{dRGE}})$ for fixed $\left< k \right>$, as shown in Fig.~\ref{Fig01}(d), 
extends to the corresponding PDFs. 

In addition, in Fig.~\ref{Fig02}:
(i) We verify that, for $\left< k \right><1$, $P(r_\mathbb{C})$ coincides with the PDF expected for
the PE: 
\begin{equation}
P_{\tbox{PE}}(r_\mathbb{C})=\mbox{const.}=1;
\label{PcPE}
\end{equation}
see left panel in Fig.~\ref{Fig02}, and also Ref.~\cite{SRP20}.
(ii) We observe, for any $\left< k \right>>1$, that $P(r_\mathbb{C})$ shows a huge peak at 
$r_\mathbb{C}\approx 1$. 
(iii) We confirm, for $\left< k \right>>7$, that $P(r_\mathbb{C})=0$ at $r_\mathbb{C}=0$, as expected
for full RMT models due to eigenvalue repulsion; see right panel in Fig.~\ref{Fig02}.

\begin{figure*}[!t]
\includegraphics[width=0.8\textwidth]{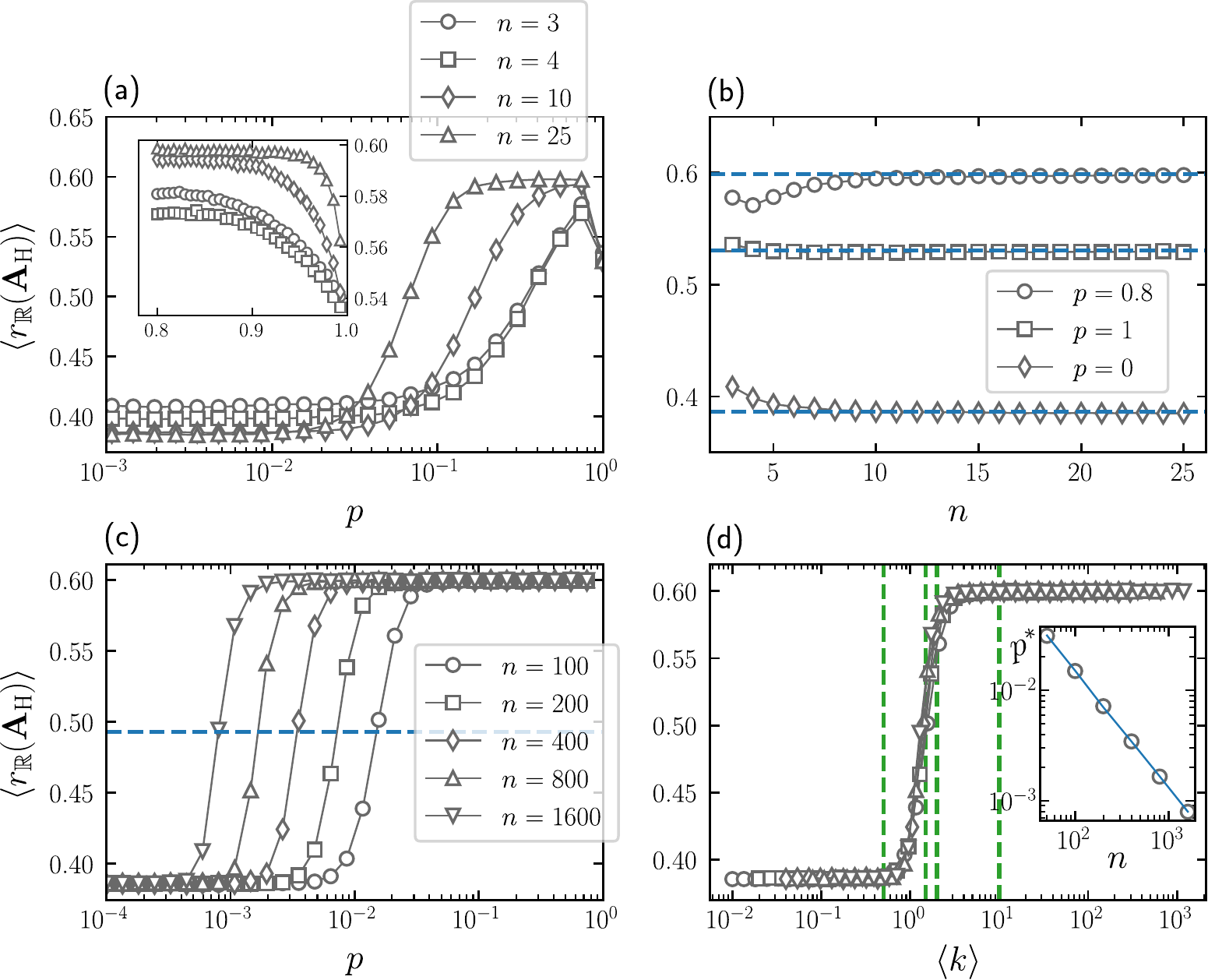}
\caption{
(a,c) Ensemble average of the ratio $r_\mathbb{R}$ for the magnetic adjacency matrices of Eq.~(\ref{AH}), 
$\left< r_\mathbb{R}(\mathbf{A}_{\tbox{M}}) \right>$, as a function of the probability $p$ for several network 
sizes $n$. The inset in panel (a) is an enlargement in the interval $p\in[0.8,1)$.
The horizontal dashed line in panel (c) indicates 
$\left< r_\mathbb{R}(\mathbf{A}_{\tbox{M}}) \right>=0.4932$.
(b) $\left< r_\mathbb{R}(\mathbf{A}_{\tbox{M}}) \right>$ as a function of $n$ for $p=0$, 0.8 and 1. 
The horizontal dashed lines at $\left< r_\mathbb{R}(\mathbf{A}_{\tbox{M}}) \right>=0.5995$, 0.5307 
and 0.3867 indicate the values of $\left< r_\mathbb{R}(\mathbf{A}_{\tbox{M}}) \right>$ for $p=0.8$, 1 and 0, respectively, at $n=1000$.
(d) Same curves of panel (c) but as a function of the average degree $\left< k \right>$.
Vertical dashed lines mark the values of $\left< k \right>$ (0.5, 1.2, 1.5, 2, and 10) chosen to report
the PDFs of $r_\mathbb{R}$, $P(r_\mathbb{R})$, in Fig.~\ref{Fig05}.
Inset: $p^*$ as a function of $n$. The dashed line is the fitting of Eq.~(\ref{scaling}) to the the data 
with fitting parameters $\mathcal{C} = 1.9563$ and $\delta = -1.0584$.
Each symbol was computed from the ratios of $10^6/n$ directed random networks $G(n,p)$.}
\label{Fig03}
\end{figure*} 
\begin{figure*}[!t]
\includegraphics[width=0.8\textwidth]{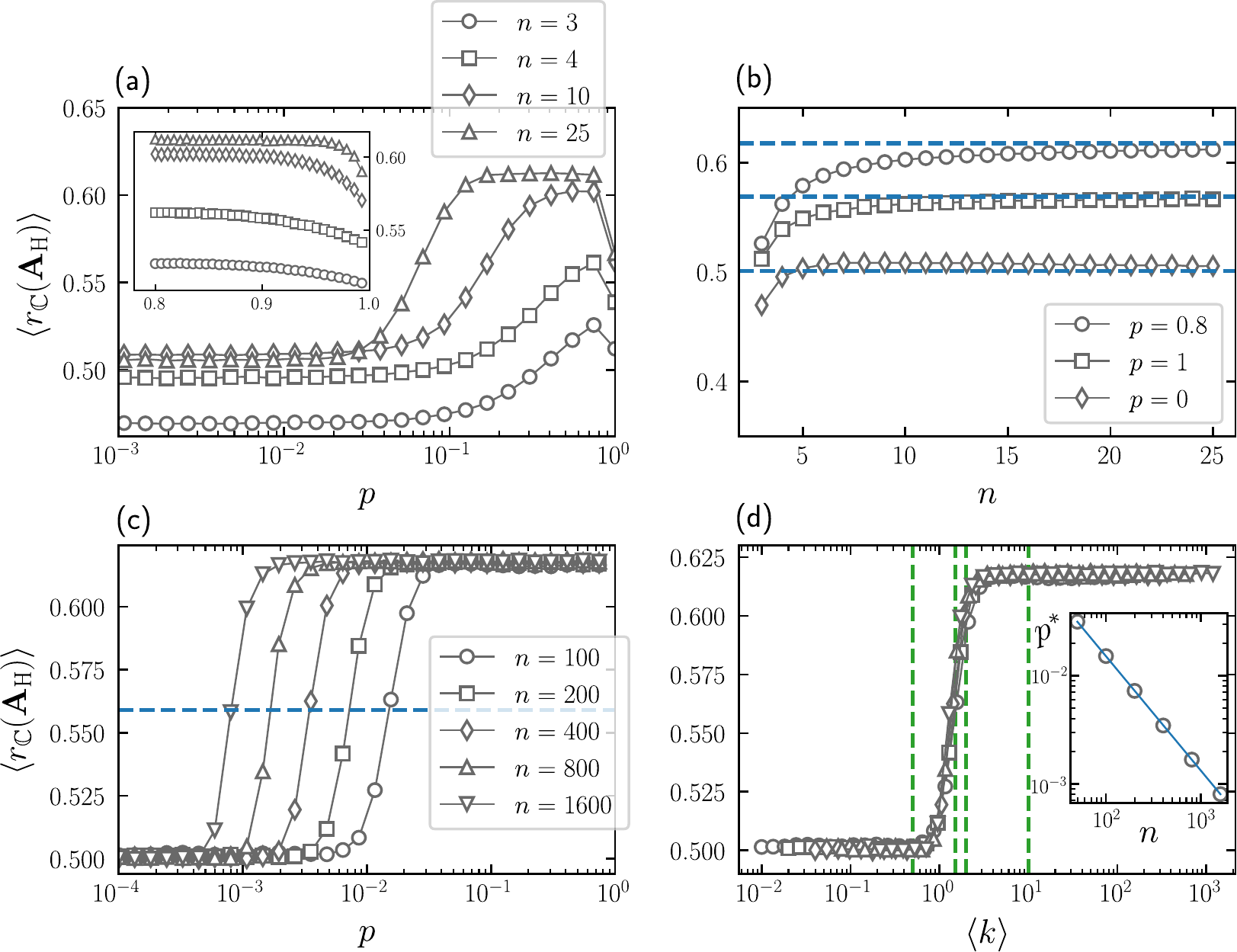}
\caption{
(a,c) Ensemble average of the ratio $r_\mathbb{C}$ for the magnetic adjacency matrices of Eq.~(\ref{AH}), 
$\left< r_\mathbb{C}(\mathbf{A}_{\tbox{M}}) \right>$, as a function of the probability $p$ for several network 
sizes $n$. The inset in panel (a) is an enlargement in the interval $p\in[0.8,1)$.
The horizontal dashed line in panel (c) indicates 
$\left< r_\mathbb{C}(\mathbf{A}_{\tbox{M}}) \right>=0.559$.
(b) $\left< r_\mathbb{C}(\mathbf{A}_{\tbox{M}}) \right>$ as a function of $n$ for $p=0$, 0.8 and 1. 
The horizontal dashed lines at $\left< r_\mathbb{R}(\mathbf{A}_{\tbox{M}}) \right>=0.6175$, 0.5688 
and 0.5006 indicate the values of $\left< r_\mathbb{R}(\mathbf{A}_{\tbox{M}}) \right>$ for $p=0.8$, 1 and 0, respectively, at $n=1000$.
(d) Same curves of panel (c) but as a function of the average degree $\left< k \right>$.
Vertical dashed lines mark the values of $\left< k \right>$ (0.5, 1.2, 1.5, 2, and 10) chosen to report
the PDFs of $r_\mathbb{C}$, $P(r_\mathbb{C})$, in Fig.~\ref{Fig05}.
Inset: $p^*$ as a function of $n$. The dashed line is the fitting of Eq.~(\ref{scaling}) to the the data 
with fitting parameters $\mathcal{C} = 1.9721$ and $\delta = -1.0581$.
Each symbol was computed from the ratios of $10^6/n$ directed random networks $G(n,p)$.}
\label{Fig04}
\end{figure*} 
\begin{figure*}[!t]
\includegraphics[width=1.0\textwidth]{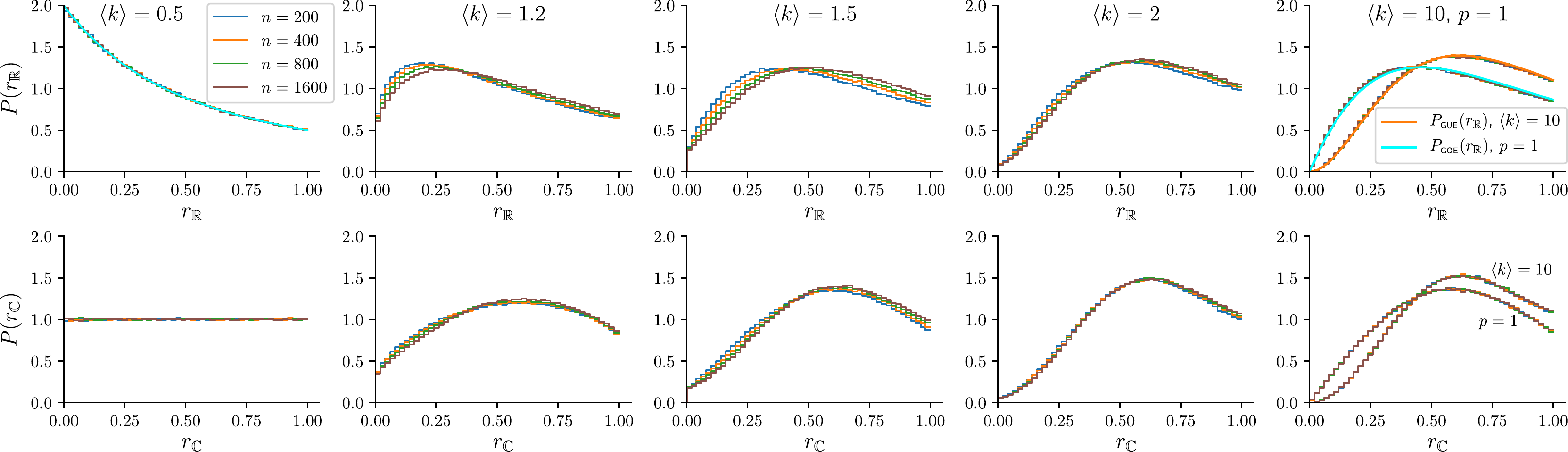}
\caption{
Probability density function of the ratios $r_\mathbb{R}$ (upper panels) and $r_\mathbb{C}$ 
(lower panels), $P(r_\mathbb{R})$ and $P(r_\mathbb{C})$, for the magnetic adjacency matrix of 
Eq.~(\ref{AH}). Each panel, corresponding to different values of the average degree $\left< k \right>$, 
contains histograms of four different network sizes $n$. The values of $\left< k \right>=0.5$, 1.2, 1.5, 
2, and 10 are marked as vertical dashed lines in Figs.~\ref{Fig03}(d) and~\ref{Fig04}(d).
In the rightmost panels the case $p=1$ is also reported. Cyan lines in the upper-left and upper-right 
panels are $P_{\tbox{PE}}(r_\mathbb{R})$ and $P_{\tbox{GOE}}(r_\mathbb{R})$, from Eqs.~(\ref{PrPE}) and (\ref{PrGOE}), respectively. The orange line in the upper-right panel is 
$P_{\tbox{GUE}}(r_\mathbb{R})$ from Eq.~(\ref{PrGUE}).
Each histogram was constructed from the ratios of $10^6/n$ directed random networks $G(n,p)$.}
\label{Fig05}
\end{figure*}

\subsection{Magnetic adjacency matrix}
\label{sec:magnetic_adj_mtrx}

Now we explore the spectral properties of the magnetic adjacency matrices $\mathbf{A}_{\tbox{M}}$.
Since $\mathbf{A}_{\tbox{M}}$ has real spectra we first use $r_\mathbb{R}$ to characterize it; later
we will also use $r_\mathbb{C}$.

In Fig.~\ref{Fig03} we show the statistics of $r_\mathbb{R}$ on $\mathbf{A}_{\tbox{M}}$. 
This figure is equivalent to Fig.~\ref{Fig01} and, in fact, it shows a very similar scenario as that reported 
for $\left< r_\mathbb{C}(\mathbf{A}_{\tbox{dRGE}}) \right>$. 
Indeed, in Fig.~\ref{Fig03} we can observe:
small-size effects mainly for $n<50$, see Figs.~\ref{Fig03}(a,b), and the scaling of
$\left< r_\mathbb{R}(\mathbf{A}_{\tbox{M}}) \right>$ with $\left< k \right>$, see 
Figs.~\ref{Fig03}(d) and Eq.~(\ref{scaling}).

Moreover, we found two important differences in the behavior of 
$\left< r_\mathbb{R}(\mathbf{A}_{\tbox{M}}) \right>$ as compared to
$\left< r_\mathbb{C}(\mathbf{A}_{\tbox{dRGE}}) \right>$. On the one hand, as expected, the curves of $\left< r_\mathbb{R}(\mathbf{A}_{\tbox{M}}) \right>$ 
show a smooth transition (in log scale) from $\approx 0.3867$ to $\approx 0.6$ when $p$ increases 
from zero to a large $p$ value, $p \approx 0.8$ in our case. However, 
$\left< r_\mathbb{R}(\mathbf{A}_{\tbox{M}}) \right>$ does not remain constant when further increasing 
$p$; instead it decreases, see the inset of Fig.~\ref{Fig03}(a), until approaching the value of $\approx 0.53$ 
at $p=1$, for large $n$.
Notice that the values of $\left< r_\mathbb{R}(\mathbf{A}_{\tbox{M}}) \right>$ reported above
(0.3867, 0.6 and 0.53; also shown in Table~\ref{T1})
correspond to those reported in~\cite{ABG13} for $\left< r_\mathbb{R}(\mbox{PE}) \right>$,
$\left< r_\mathbb{R}(\mbox{GUE}) \right>$ and $\left< r_\mathbb{R}(\mbox{GOE}) \right>$, respectively.
Here, GUE stands for a RMT ensemble known as the Gaussian Unitary Ensemble which is formed 
by Hermitian random $n\times n$ matrices where the real and imaginary parts of their complex
entries are independent and identically distributed Gaussian variables.
Therefore, we observe that the spectral statistics of $\mathbf{A}_{\tbox{M}}$ transits first from PE to GUE 
statistics and later from GUE to GOE statistics.
This triple transition (PE-to-GUE-to-GOE) can be understood from the definition of $\mathbf{A}_{\tbox{M}}$ 
itself, see Eq.~(\ref{AH}): Clearly, when $p\to 0$, $\mathbf{A}_{\tbox{M}}$ becomes an almost-diagonal 
real random matrix, so its spectral statistics is expected to be close to the PE statistics.
Then, for intermediate values of $p$ most of the off diagonal entries are imaginary, so we observe
clear GUE-like statistics even though the matrix $\mathbf{A}_{\tbox{M}}$ is far from being a member
of the GUE. We numerically found that the GUE characteristics appear in the parameter range from 
$\left< k \right>\approx 4$ to $p\approx 0.8$ (for large $n$).
At $p=1$ the number of imaginary entries of $\mathbf{A}_{\tbox{M}}$ becomes zero, so its spectral 
statistics is expected to be close to the GOE statistics, even when 
$\mathbf{A}_{\tbox{M}}$ is not strictly a member of the GOE.
It is important to stress that in the GUE-to-GOE transition regime, the curves of 
$\left< r_\mathbb{R}(\mathbf{A}_{\tbox{M}}) \right>$ do not scale with $\left< k \right>$; so we are 
avoiding this regime in Figs.~\ref{Fig03}(c,d).

\setlength{\tabcolsep}{6pt}
\begin{table}[b!]
\caption{Reference average values of the ratios $r_\mathbb{R}$ and $r_\mathbb{C}$ for the random 
adjacency matrices used in this work. To compute the averages, the spectra of $10^3$ adjacency 
matrices of size $n=1000$ were used; i.e.~approx.~$10^6$ ratios were used to compute the averages.}
\label{T1}
\begin{tabular*}{\columnwidth}{ c | c | c | c  | c }  
\hline
 & PE & $\mathbf{A}_{\tbox{dRGE}}(p=1)$ & $\mathbf{A}_{\tbox{M}}(\langle k \rangle=10)$  & $\mathbf{A}_{\tbox{M}}(p=1)$ \\
\hline
$\left\langle r_\mathbb{C} \right\rangle$ 
& 0.5006 & 0.7370 & 0.6175 & 0.5688 \\
$\left\langle r_\mathbb{R} \right\rangle$  
& 0.3867 & -- & 0.5995 & 0.5307 \\
\hline
\end{tabular*}
\end{table}

On the other hand, the PE-to-GOE transition regime of $\left< r_\mathbb{R}(\mathbf{A}_{\tbox{M}}) \right>$, 
starting at $\left< k \right>\approx 0.7$, is slightly narrower than the PE-to-RGE transition regime of 
$\left< r_\mathbb{C}(\mathbf{A}_{\tbox{dRGE}}) \right>$. Here, the transition regime is observed for
$0.7<\left< k \right><4$.

In addition, to complete the characterization of the spectra of $\mathbf{A}_{\tbox{M}}$, in 
Fig.~\ref{Fig04} we present the statistics of $r_\mathbb{C}$, that can also be computed for real spectra.
It is remarkable to note that that $\left< r_\mathbb{C}(\mathbf{A}_{\tbox{M}}) \right>$ provides equivalent 
information than $\left< r_\mathbb{R}(\mathbf{A}_{\tbox{M}}) \right>$, as can be seen by comparing
Figs.~\ref{Fig03} and~\ref{Fig04}. In particular, in Fig.~\ref{Fig04} we observe:
small-size effects mainly for $n<50$, see Figs.~\ref{Fig04}(a,b); 
the scaling of $\left< r_\mathbb{C}(\mathbf{A}_{\tbox{M}}) \right>$ with $\left< k \right>$, see 
Fig.~\ref{Fig04}(d);
the triple transition PE-to-GUE-to-GOE, see Fig.~\ref{Fig04}(a); and 
the PE-to-RGE transition regime in the interval $0.7<\left< k \right><4$, see Fig.~\ref{Fig04}(d).
In Table~\ref{T1} we report the asymptotic values $\left< r_\mathbb{C}(\mbox{PE}) \right>$,
$\left< r_\mathbb{C}(\mbox{GUE}) \right>$ and $\left< r_\mathbb{C}(\mbox{GOE}) \right>$,
that (as far as we know) were not reported before.

Finally, in Fig.~\ref{Fig05} we show the PDFs for the ratios $r_\mathbb{R}$ and $r_\mathbb{C}$, 
$P(r_\mathbb{R})$ and $P(r_\mathbb{C})$, respectively, for the magnetic adjacency matrix 
$\mathbf{A}_{\tbox{M}}$ at representative values of $\left< k \right>$.
Indeed, with this figure we verify the invariance of $P(r_\mathbb{R})$ and $P(r_\mathbb{C})$ for
fixed $\left< k \right>$, with clear small-size effects for intermediate values of $\left< k \right>$. 
Moreover, we also validate the PE-to-GUE-to-GOE transition observed for
$\left< r_\mathbb{R}(\mathbf{A}_{\tbox{M}}) \right>$ in Fig.~\ref{Fig03}.
Note that:
when $\left< k \right>< 1$, $P(r_\mathbb{R})$ is well reproduced by the prediction for the
PE (see the cyan curve in the upper-left panel) which is given by ~\cite{ABG13}
\begin{equation}
P_{\tbox{PE}}(r_\mathbb{R}) = \frac{2}{(1+r_\mathbb{R})^2}.
\label{PrPE}
\end{equation}
In the parameter range from $\left< k \right>\approx 7$ to $p\approx 0.8$ (for large $n$),
the $P(r_\mathbb{R})$ coincides with the prediction for the GUE~\cite{ABG13} 
\begin{equation}
P_{\tbox{GUE}}(r_\mathbb{R}) = \frac{81\sqrt{3}}{2\pi} \frac{(r_\mathbb{R}+r_\mathbb{R}^2)^2}{(1+r_\mathbb{R}+r_\mathbb{R}^2)^4} \ ,
\label{PrGUE}
\end{equation}
see the orange curve in the upper-right panel; while for $p=1$, $P(r_\mathbb{R})$ corresponds to
the prediction for the GOE~\cite{ABG13}
\begin{equation}
P_{\tbox{GOE}}(r_\mathbb{R}) = \frac{27}{4} \frac{r_\mathbb{R}+r_\mathbb{R}^2}{(1+r_\mathbb{R}+r_\mathbb{R}^2)^{5/2}} \ ,
\label{PrGOE}
\end{equation}
see the cyan curve in the upper-right panel.

In the case of $r_\mathbb{C}$, we can only compare its PDF with $P_{\tbox{PE}}(r_\mathbb{C})$, 
see Eq.~(\ref{PcPE}), which indeed reproduces well the $P(r_\mathbb{C})$ of $\mathbf{A}_{\tbox{M}}$ 
when $\left< k \right>< 1$; see the lower-left panel of Fig.~\ref{Fig05}. We note that (as far as we know) exact expressions for $P_{\tbox{GUE}}(r_\mathbb{C})$
and $P_{\tbox{GOE}}(r_\mathbb{C})$ are not known.
We also confirm, for $\left< k \right>>7$, that both $P(r_\mathbb{R})=0$ at $r_\mathbb{R}=0$ and 
$P(r_\mathbb{C})=0$ at $r_\mathbb{C}=0$, as usual in full RMT models; see the right panels of 
Fig.~\ref{Fig05}.

\color{black}

\begin{figure}[!t]
\includegraphics[width=1.0\columnwidth]{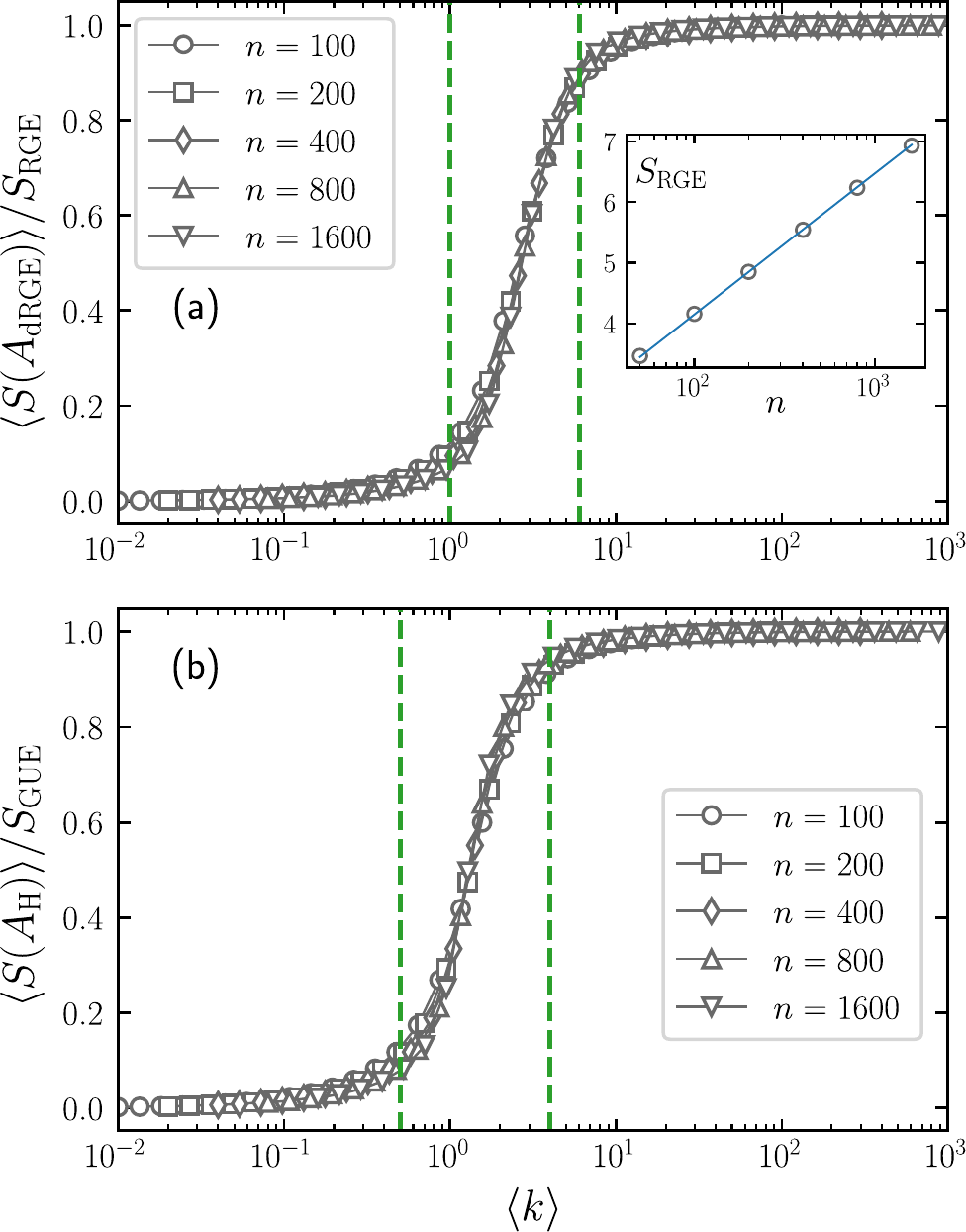}
\caption{
Normalized average Shannon entropy $\bra S \ket$ as a function of the average degree 
$\left< k \right>$ for the adjacency matrices (a) $\mathbf{A}_{\tbox{dRGE}}$ and (b) 
$\mathbf{A}_{\tbox{M}}$, corresponding to directed networks of size $n$. 
In (a) [(b)] we are normalizing $\bra S(\mathbf{A}_{\tbox{dRGE}}) \ket$ 
[$\bra S(\mathbf{A}_{\tbox{M}}) \ket$] to $S_{\tbox{RGE}}$ [$S_{\tbox{GUE}}$].
The inset in (a) shows the numerically computed $S_{\tbox{RGE}}$ as a function of $n$. 
The dashed line is a fitting to the data that provides $S_{\tbox{RGE}}\approx \ln(n/1.56)$.
Vertical dashed lines indicate the transition regime as deduced from $\bra r_\mathbb{C} \ket$:
(a) $1<\left< k \right><7$ and (b) $0.7<\left< k \right><4$.
Each symbol was computed by averaging over $10^6$ eigenvectors.}
\label{Fig06}
\end{figure}

\section{Delocalization transition}

In the previos Section we characterized the PE-to-RGE transition of $\mathbf{A}_{\tbox{dRGE}}$ and 
the PE-to-GUE transition of $\mathbf{A}_{\tbox{M}}$ by means of their spectral properties. These 
transitions, indeed, imply to localization-to-delocalization transition (or simply known as delocalization transition)
of the corresponding eigenvectors; i.e.~the eigenvectors should go from localized (in the PE regime) 
to extended (in the RGE or GUE regimes). Thus, in the following we verify this statement.

To measure quantitatively the spreading of eigenvectors in a given basis, i.e., their localization 
properties, the information or Shannon entropy $S$ is commonly used~\cite{MK98}. Moreover, 
it has been widely used to characterize the eigenvectors of the adjacency matrices of random 
network models. For the eigenvector $\Psi^k$, associated with the eigenvalue $\lambda_k$, 
$S$ is given as 
\begin{equation}
S^k = -\sum_{l = 1}^n \mid \Psi_l^k \mid^2 \ln \mid \Psi_l^k \mid^2 .
\label{S}
\end{equation}
This measure provides the number of main components of the eigenvector $\Psi^k$. 

We average over all eigenvectors of ensembles of adjacency matrices $\mathbf{A}_{\tbox{dRGE}}$ 
and $\mathbf{A}_{\tbox{M}}$ to compute $\langle S \rangle$, such that for each combination 
$(n,p)$ we use $10^6$ eigenvectors. With definition (\ref{S}), when $p \to 0$, since the eigenvectors of 
$\mathbf{A}_{\tbox{dRGE}}$ and $\mathbf{A}_{\tbox{M}}$ have only one main component with magnitude 
close to one, $\langle S \rangle\approx 0$. On the other hand, for $p\to 1$, the fully chaotic eigenvectors 
extend over the $n$ available vertices of the directed network, so~\cite{MK98} 
$\langle S \rangle \approx \ln (n) - {\cal C}$, for large $n$; where ${\cal C}$ is a constant (independent 
of $n$) specified by the symmetries of a given random matrix ensemble.

For any network size $n$, $\bra S \ket$ displays a similar functional form as a function of $p$: 
The curves of $\bra S \ket$ show a smooth transition from approximately zero to 
$S_{\tbox{MAX}}$ when $p$ increases from $p\sim 0$ (mostly isolated vertices) to one (complete 
graphs). Recall that when $\bra S \ket \approx 0$ the corresponding eigenvectors are localized 
(i.e.,~$\bra S \ket \approx 0$ defines the localized regime). In contrast, when 
$\bra S \ket \approx S_{\tbox{MAX}}$, the corresponding eigenvectors are delocalized. 
Thus, the curves of $\bra S \ket$ versus $p$ indicate the delocalization transition of the eigenvectors
of our random network model.
In the case of $\mathbf{A}_{\tbox{dRGE}}$, $S_{\tbox{MAX}}=S_{\tbox{RGE}}$; that is, 
$S_{\tbox{MAX}}$ corresponds to the Shannon entropy of the eigenvectors of the RGE. 
Moreover, since we do not have an explicit expression for $S_{\tbox{RGE}}$ we compute
it numerically for the network sizes used in this work, see the inset of Fig.~\ref{Fig06}(a), and found
that $S_{\tbox{RGE}}\approx \ln(n/1.56)$.
For $\mathbf{A}_{\tbox{M}}$, $S_{\tbox{MAX}}=S_{\tbox{GUE}}\approx  \ln(n/1.53)$~\cite{MK98}.
Therefore, in Fig.~\ref{Fig06} we present the normalized average Shannon entropy 
$\bra S \ket/S_{\tbox{MAX}}$ already as a function of the average degree $\left< k \right>$ 
(i.e.~after the scaling analysis of the previous Section) for directed random networks represented 
by the matrices $\mathbf{A}_{\tbox{dRGE}}$ and $\mathbf{A}_{\tbox{M}}$. 
From this figure we clearly observe that the curves of $\bra S \ket/S_{\tbox{MAX}}$ 
(i) demonstrate the delocalization transition of the eigenvectors of both $\mathbf{A}_{\tbox{dRGE}}$ 
and $\mathbf{A}_{\tbox{M}}$, as anticipated, and
(ii) scale with $\left< k \right>$, as expected. Finally, we note that very recently Metz and Neri~\cite{metz2020localization} have put forward calculations on the delocalization-localization transition of random directed networks. The authors showed analytically that the eigenvectors related to the largest eigenvalue and the eigenvalue at the boundary of the spectral bulk go from a localized to a delocalized regime as the connectivity is increased, which is in agreement with Fig.~\ref{Fig06}.

\section{Summary and Conclusions}

In this work we have used real and complex spacing ratio measures to characterize the spectra of directed random networks. The great advantage of the spacing ratio approach over the traditional characterization via level-spacing distributions, $P(s)$, is that the former does not require any unfolding procedure -- a task that, by contrast, usually depends on a prior knowledge of the spectral density, and whose calculation in some situations is numerically unfeasible. However, it is fair to mention that, spectral properties of directed networks have been successfully studied by the use of $P(s)$, see e.g.~\cite{BLXG15}.

We have investigated two \textcolor{black}{adjacency matrix representations of Erd\H{o}-R\`enyi (ER) random networks: a diluted version of the real Ginibre ensemble (dRGE), i.e.,~sparse non-Hermitian random matrices, and an operator defined in Refs.~\cite{guo2017hermitian,liu2015hermitian} leading to sparse Hermitian random matrices.} For the first ensemble, which yields complex spectra, we computed the complex spacing ratio $r_{\mathbb{C}}$, introduced recently in Ref.~\cite{SRP20}, which is defined as the ratio between the distance of the nearest neighbor eigenvalue over the distance to the next-to-nearest-neighbor one. We have shown that the average measure, $\langle r_{\mathbb{C}}\rangle$,
undergoes a smooth transition \textcolor{black}{from Poisson to Ginibre statistics} as a function of the network connectivity; this transition was verified to occur at lower probabilities upon the increase of the network size, thus suggesting the existence of a scaling parameter relating networks \textcolor{black}{with different parameter combinations}. In effect, by scaling $\langle r_{\mathbb{C}}\rangle$ in terms of the average degree $\langle k \rangle$, we found that the curves \textcolor{black}{$\langle r_{\mathbb{C}}\rangle$ vs.~$\langle k \rangle$} corresponding \textcolor{black}{to networks of} different sizes collapse onto a universal curve, in consonance with the universal properties of undirected ER networks~\cite{GAC18,MM19,TFM19}. From the universal transition curve we have identified three distinct statistical regimes: For $\langle k \rangle < 1$, i.e.~below the percolation threshold, $\langle r_{\mathbb{C}}\rangle \approx 0.5$, which coincides with the value of  $\langle r_{\mathbb{C}}\rangle$ for the Poisson ensemble (PE) of RMT. For denser networks, with $\langle k \rangle > 7$, one obtains the corresponding value of the real Ginibre ensemble (RGE), that is $\langle r_{\mathbb{C}}\rangle \approx 0.737$. The range $1 < \langle k \rangle < 7$ defines then the intermediate region in the transition from PE to RGE statistics.     

Although complex spacing ratios have been conceived for the analysis of complex spectra, they can also be applied to the characterization of real spectra. We exemplified this when studying, in Sec.~\ref{sec:magnetic_adj_mtrx}, the \textcolor{black}{{\it magnetic}} Hermitian matrices \textcolor{black}{obtained from Eq.~(\ref{AH})}. In fact, we have shown that $\langle r_{\mathbb{C}} \rangle$ provides equivalent information \textcolor{black}{than the average real} ratio $\langle r_{\mathbb{R}} \rangle$ [see Figs.~\ref{Fig03} and~\ref{Fig04}]; that is, both measures display \textcolor{black}{a smooth delocalization transition as a function of the connection probability $p$, which becomes universal under the} scaling with $\langle k \rangle$. Comparing the dRGE studied in Sec.~\ref{sec:diluted_real_ginibre_ensemble} with the magnetic matrices of Sec.~\ref{sec:magnetic_adj_mtrx}, we have seen that both ensembles exhibit qualitatively a similar evolution of $\langle r_{\mathbb{C}} \rangle$ with respect to the network connectivity, except for values of $p$ close to 1: As the this \textcolor{black}{limit} is approached, both $\langle r_{\mathbb{C}} \rangle$ and $\langle r_{\mathbb{R}}  \rangle$, \textcolor{black}{on the magnetic matrices}, decay smoothly. The reason for this effect resides in the very definition of the magnetic matrices in Eq.~(\ref{AH}): For $p=1$, the imaginary \textcolor{black}{entries vanish and the magnetic} ensemble becomes equivalent to the Gaussian Orthogonal Ensemble (GOE). Therefore, as the average connectivity is increased, the spectrum of the magnetic matrices defined in Eq.~(\ref{AH}) \textcolor{black}{transits from PE to GOE statistics, and subsequently to GUE statistics.}
 
Prior studies on real and complex spacing ratios \textcolor{black}{arising from Hermitian and non-Hermitian systems, respectively, have shown that such measures are able to distinguish between integrable and chaotic spectra, see e.g.~\cite{SRP20,CR20,CDK14,SKK20}}. Here we showed that the same quantities can also  differentiate the disconnected phase ($\langle k \rangle < 1$), in which the \textcolor{black}{directed} network is divided into several small components, and the connected phase ($\langle k \rangle > 1$), where a giant component connecting the majority of the nodes \textcolor{black}{emerges}. In this context, average spacing ratios could serve as universal indicators  to define sparse and dense connectivity regimes for undirected and directed networks. For instance, it is known that mean-field calculations for dynamical processes on networks perform well for ``sufficiently dense'' structures~\cite{gleeson2012accuracy}; however, precise bounds for the accuracy of such approximations have not yet been established. Thus, it would be interesting to relate delocalization transitions, as quantified by spacing ratios, with transitions \textcolor{black}{associated to dynamical processes (such as epidemic spreading and synchronization) in order to quantify accurately the limits of mean-field approximations in terms of spectral measurements.}
It would also be pertinent to extend the analysis performed here
to systems with more heterogeneous degree distributions, such as scale-free networks. We leave these open issues for future works.   





\appendix

\section{Relation between the magnetic operator and the Hermitian adjacency operator of Refs.~\cite{guo2017hermitian,liu2015hermitian}}
\label{appendixMoharMagnetic}

Here we show that the Hermitian adjacency matrix recently introduced and studied in Refs.~\cite{guo2017hermitian,liu2015hermitian}
is, in fact, a special case of the magnetic operator defined in~\cite{lieb1993,berkolaiko2013,fanuel2017magnetic}.

Let $G(V, E)$ be an unweighted directed graph, where $V$ is the set of vertices and $E=\{(u, v)| u, v \in V\}$ 
is the set of edges. The adjacency operator of Refs.~\cite{guo2017hermitian,liu2015hermitian} can be defined as an unweighted 
directed graph $\mathcal H:V\times V\rightarrow \mathbb C$ whose adjacency matrix is given by
\begin{align}
\mathcal H(u, v) = \begin{cases}
1, \ \ \textrm{ if } (u, v) \in E \textrm{ and } (v, u) \in E\\
i, \ \ \textrm{ if } (u, v) \in E \textrm{ and } (v, u) \notin E\\
-i, \ \ \textrm{ if } (u, v) \notin E \textrm{ and } (v, u) \in E\\
0, \  \ \textrm{otherwise}.
\end{cases}
\end{align}
Let a weight function $\mathcal W: V\times V\rightarrow \{0, 1\}$ such that
\begin{align}
\mathcal W(u, v) =\begin{cases}
1, \textrm{ if }(u,v) \in E\\
0, \textrm{ otherwise}.
\end{cases}
\end{align}
The magnetic adjacency operator is given by
\begin{align}
\mathcal M(u, v) =
\frac{\mathcal W(u, v)+\mathcal W(v, u)}{2}
	e^{
    \mathrm{i} \phi \mathcal A(u,v) 
    },
\end{align}
where $ \mathcal A(u, v) = \mathcal W(v,u)-\mathcal W(u,v)$. For $\phi = -\frac{\pi}{2}$ we have
\begin{align}
\mathcal M(u, v) = \begin{cases}
1, \ \ \textrm{ if } (u, v) \in E \textrm{ and } (v, u) \in E\\
\frac{\mathrm i}{2}, \ \ \textrm{ if } (u, v) \in E \textrm{ and } (v, u) \notin E\\
\frac{-\mathrm i}{2}, \ \ \textrm{ if } (u, v) \notin E \textrm{ and } (v, u) \in E\\
0, \  \ \textrm{otherwise},
\end{cases}.
\end{align}
which is very close to $\mathcal H(u, v)$. Moreover, we can recover the operator $\mathcal H(u, v)$, 
exactly (without the factor $1/2$), by the use of the weight function
\begin{align}
\mathcal W(u, v) =\begin{cases}
1, \textrm{ if }(u,v) , (v, u) \in E\\
2 , \textrm{ if }(u,v)  \in E \textrm { and } (v, u) \notin E\\
0, \textrm{ otherwise}
\end{cases}
\end{align}
and setting $\phi = -\frac{\pi}{4}$. 

The magnetic adjacency operator and, consequently, the magnetic Laplacian operator, was proposed  
by Lieb and Loss~\cite{lieb1993} when studying the problem of a quantum particle in a discrete space.  
Recently, this magnetic operator emerged as an important tool in the study of mathematical properties 
of graphs~\cite{berkolaiko2013} and in the development of algorithms for directed networks such as 
community detection~\cite{fanuel2017magnetic}, signal processing~\cite{signalMagnetic2020} and 
network characterization~\cite{resende2020characterization}. Indeed, the operator $\mathcal{M}$ can be applied to more general graphs than 
the Hermitian adjacency matrix $\mathcal H(u, v)$. Interestingly, the equivalence between 
$\mathcal H(u, v)$ and the magnetic Laplacian operators has remained, to our knowledge, unnoticed in previous works.

\begin{acknowledgments}
T.P. acknowledges FAPESP (Grants No. 2016/23827-6). BM thanks CAPES for financial support. FAR acknowledges the Leverhulme Trust, CNPq (Grant No. 305940/2010-4) and FAPESP (Grants No. 2016/25682-5 and grants 2013/07375-0) for the financial support given to this research. J.A.M.-B. acknowledges financial support from FAPESP (Grant No.~2019/ 06931-2), Brazil, 
CONACyT (Grant No.~2019-000009-01EXTV-00067) and PRODEP-SEP (Grant No.~511-6/2019.-11821), Mexico. Luciano da F. Costa
thanks CNPq (grant no. 307085/2018-0) and NAP-PRPUSP for sponsorship.
\end{acknowledgments}

\bibliography{bibliography}

\end{document}